\def \cm{~\rm{cm}}
\def \s{~\rm{s}}
\def \km{~\rm{km}}

\def \AU{~\rm{AU}}
\def \erg{~\rm{erg}}

\def \yr{~\rm{yr}}

\def \kpc{~\rm{kpc}}

\documentclass[12pt,preprint]{aastex}

\usepackage{natbib}
\usepackage{epsfig}
\begin{document}

\title{BUBBLES IN PLANETARY NEBULAE AND CLUSTERS OF GALAXIES:
JET BENDING}

\author{Noam Soker and Gili Bisker}
\affil{Department of Physics, Technion$-$Israel
Institute of Technology, Haifa 32000 Israel;
soker@physics.technion.ac.il.}

\begin{abstract}
We study the bending of jets in binary stellar systems.
A compact companion accretes mass from the slow wind of the mass-losing
primary star, forms an accretion disk, and blows two opposite jets.
These fast jets are bent by the slow wind.
Disregarding the orbital motion, we find the dependence of the bending angle
on the properties of the slow wind and the jets.
Bending of jets is observed in planetary nebulae which are thought to be
the descendants of interacting binary stars.
For example, in some of these planetary nebulae the two bubbles (lobes)
which are inflated by the two opposite jets, are displaced to the same side
of the symmetry axis of the nebula.
Similar displacements are observed in bubble pairs in the center of some
clusters and groups of galaxies.
We compare the bending  of jets in binary stellar systems with that in clusters of
galaxies.

\end{abstract}


\section{INTRODUCTION}

The nebular gas in planetary nebulae (PNs) originates in the envelope of asymptotic
giant branch (AGB) stars that are the descendants of intermediate mass
stars (initial masses $\sim 1-8 M_\odot$).
Such stars rotate very slowly, and their mass loss is expected to be spherical.
Indeed, AGB stellar winds usually consist of a more or less spherically
symmetric outflow at rates of $\sim 10^{-7}-10^{-5} M_\odot \yr ^{-1}$.
Most PNs, though, possess a global axisymmetrical structure rather than a spherical
structure in their inner region, indicating a non-spherical shaping process.
Among the several PN shaping models (Balick \& Frank 2002), one
of the most successful is the jet-shaping model.
If the jets are not well collimated they are termed
collimated fast wind (CFW).
The presence of jets in PNs was deduced from observations
more than 20 years ago (e.g., Feibelman 1985).
Gieseking et al. (1985) found collimated outflow in the PN NGC 2392,
and noted the similarity of these jets with that of young stellar objects,
and speculated that such outflows exist in many similar PNs.
On the theoretical side, Morris (1987) suggested that two jets
blown by an accreting companion (the secondary star) can form bipolar nebulae.
This model is strongly supported by the similarity of bipolar PNs to
many bipolar symbiotic nebulae which are known to be shaped by jets
(e.g., Schwarz et al. 1989; Corradi \& Schwarz 1995).
Soker (1990) proposed that the two fast low-ionization emission blobs
(FLIERs or {\it ansae}) along the symmetry axis of many elliptical PNs are formed
by jets blown during the last phase of the AGB or the post-AGB phase of
the PN progenitor.
The high quality HST images led Sahai \& Trauger (1998) to suggest that
in many PNs the non-spherical structures are formed solely by jets.

Projecting from similar astronomical objects, the formation of massive jets,
to distinguish from magnetized low density pulsar jets,  require
the presence of accretion disks.
The only source of angular momentum sufficient to form accretion disks in evolved
stars is the orbital angular momentum of a stellar (or in some cases substellar)
companion.
The disk can be formed around the progenitor during the late post-AGB phase,
when it is already small (Soker \& Livio 1994), or, more likely, around a
stellar companion accreting mass, forming an accretion disk, and blowing two jets.

The past seven years have seen further consolidation of the bipolar jet-shaping
model in binary systems, addressed both in observations
(e.g., Parthasarathy et al. 2000; Sahai \& Nyman 2000; Miranda et al. 2001a,b;
Corradi et al. 2001; Guerrero et al. 2001; Vinkovic et al. 2004; Huggins et al. 2004;
Pena et al.\ 2004; Balick \& Hajian 2004; Arrieta et al. 2005;
Oppenheimer et al. 2005; Sahai et al. 2005)
and in theory (e.g., Soker 2002, 2005; Lee \& Sahai 2003, 2004; Livio \& Soker 2001;
Garcia-Arredondo\& Frank 2004; Velazquez et al. 2004; Riera et al. 2005).
Many of the PNs in the observations listed above posses point symmetric morphology,
i.e., several symmetry axes rotate with respect to each other through a common origin,
indicating precessing jets.
The most likely explanation for precession is an accretion disk in the presence of
a companion.
Soker \& Rappaport (2000) further discussed the jet shaping process and have shown
that the statistical distribution of bipolar PNs can be accounted for in
the binary model.
Further support for the formation of jets in binary systems comes from X-ray
observations hinting at jets in a PN (Kastner et al. 2003) similar to X-ray jets in symbiotic systems
(Kellogg et al. 2001; Galloway \& Sokoloski 2004).
Garcia-Arredondo \& Frank (2004) were the first to conduct 3D numerical simulations
of the interaction of jets launched by a secondary star with the slow primary wind.
Their high quality results strengthen the general stellar-binary jets model, and in
particular the conjecture (Soker \& Rappaport 2000) that a narrow waist can be formed
by jets.
It should be stressed that not all PNs are shaped by jets, but bubble pairs are
formed by jets.
X-ray images of active galactic nuclei in clusters of galaxies indeed show
that double-jets, observed in the radio band, can form a bubble pair with
a narrow waist between them, similar to narrow waists in PNs with no need
for enhanced equatorial mass loss rate,
although enhanced equatorial mass loss rate might occur in many PNs.

The subject of the similarity between some morphological structures in
clusters of galaxies, as revealed via X-ray observations, and in PNs,
as revealed in the visible band, was studied in a series of four papers.
\newline
{\it Paper 1} ( Soker 2003b).
In that paper ( see also Soker 2003a, and section 5 in Soker 2004c)
the similarity in morphological structures was discussed
\footnote{The similar morphologies are compared in the appendix
of the astro-ph version of the present paper.}.
This similarity is not trivial. Two opposite jets are observed in many young stellar
objects (YSOs), however, bubbles pairs similar to those in PNs and in
clusters of galaxies are not usually observed around YSOs.
\newline
{\it Paper-2 } (Soker 2004a).
It was found that to inflate fat, more or less spherical, bubbles the
opening angle of the jets should be large; the half opening
angle measured from the symmetry axis of each jet should typically be
$\alpha \gtrsim 40 ^\circ$, or the jets might precess.
\newline
{\it Paper-3} (Soker 2004b).
Paper 3 studies the stability of off-center low-density fat bubbles
in clusters of galaxies and in PNs to the Rayleigh-Taylor instability.
\newline
{\it Paper 4} (Pizzolato \& Soker 2005).
Pizzolato \& Soker examined the point symmetric structure
of the bubble pair in the cluster MS 0735.6+7421 (McNamara et al. 2005) and
compared it to the point symmetric structure of PNs.
Point symmetric PNe are thought to be shaped by stellar binary interactions;
namely, the presence of a companion to the PN's progenitor star is required.
Pizzolato \& Soker (2005) suggested that similar point-symmetric structures
in the X-ray deficient cavities of galaxy clusters might be associated with
the presence of massive binary black holes.

In this paper, the fifth in the series, we examine the bending of the
two jets, and the subsequent bending of the two bubbles inflated by the jets,
to the same side of their original symmetry axis (jets' axis).
Such displacement relative to the symmetry axis of bubbles touching the center
is seen, for example, in the Perseus cluster of galaxies (Fabian et al. 2000;
Dunn et al. 2006), and in the PN NGC 3587 (Guerrero et al. 2003).
Dunn et al. (2006) discuss the departure of the two bubbles from their alignment
along a cluster center and explain this departure by the two opposite bubbles
detaching from the precessing jets at different times.
We consider this displacement to result from the ram pressure of
the intra cluster medium (ICM).
Displacement of bubbles at a distance from the center are seen
in the PN NGC 6886 (Terzian \& Hajian 2000), and the group of galaxies
HCG 62 (Vrtilek et al. 2002).
We focus on PNs and related binary stellar objects, e.g.,
the massive binary stellar system $\eta$ Carinae ($\S 2.1$).
The departure of PNs and related binary systems from axisymmetry has
been previously studied (Soker \& Hadar 2002 and references therein).
Our goal here is to derive a simple expression for the bending of jets
in binary stellar systems ($\S 2.2$).
This expression is not a substitute for future numerical simulations.
The results for typical binary systems ($\S 2.2$) can account for some
morphological structures in PNs and related systems.
Readers interested in only using the relations and the results, can skip
$\S 2.1$ and go directly to $\S 2.2$.
In $\S 3$ we compare the situation with jet bending in cooling
flow clusters, and $\S 4$ is a summary.

\section{BENDING IN A BINARY STELLAR SYSTEM}
\subsection{Assumptions and Equations}

When a compact secondary star accretes from the AGB (or post-AGB) stellar wind
only part of the AGB wind is accreted, and the rest expands outward and forms the
medium that the jets expand into.
In addition, when the jet is still close to the binary system, the AGB wind hits
the jet on its side, causing the jet to deflect (Soker \& Rappaport 2000).
Like precession, this can have large effects on the descendant PN morphology.
However, while precession leads to point-symmetric nebula, the deflection of the two
oppositely ejected jets is to the same side, leading the two opposite lobes
to be bent to the same side; this is the {\it bent} departure from axisymmetry
according to the classification of Soker \& Hadar (2002).
The bending interaction can clear the way to radiation, possibly ionizing radiation,
from the central binary system to more strongly illuminate
the same side in both lobes (bubbles).
Due to the orbital motion, this structure forms a revolving light source.
Livio \& Soker (2001) suggested such a revolving ionizing source model to explain the
positional shift of the bright knots in the inner nebular lobes of the M2-9 nebula
(Doyle et al. 2000).
Soker \& Rappaport (2000) derived a simple expression for the bending angle
of a narrow jet.
In this section we relax some of the assumptions made by Soker \& Rappaport and
derive a more accurate expression for the bending angle, while still keeping the
expression simple.
The goal is to derive a simple approximate relation that will give the jet's bending
angle upon specifying the jet's parameters and slow wind parameters.

\begin{figure}
\vskip -2.2 cm
\includegraphics[width =150mm]{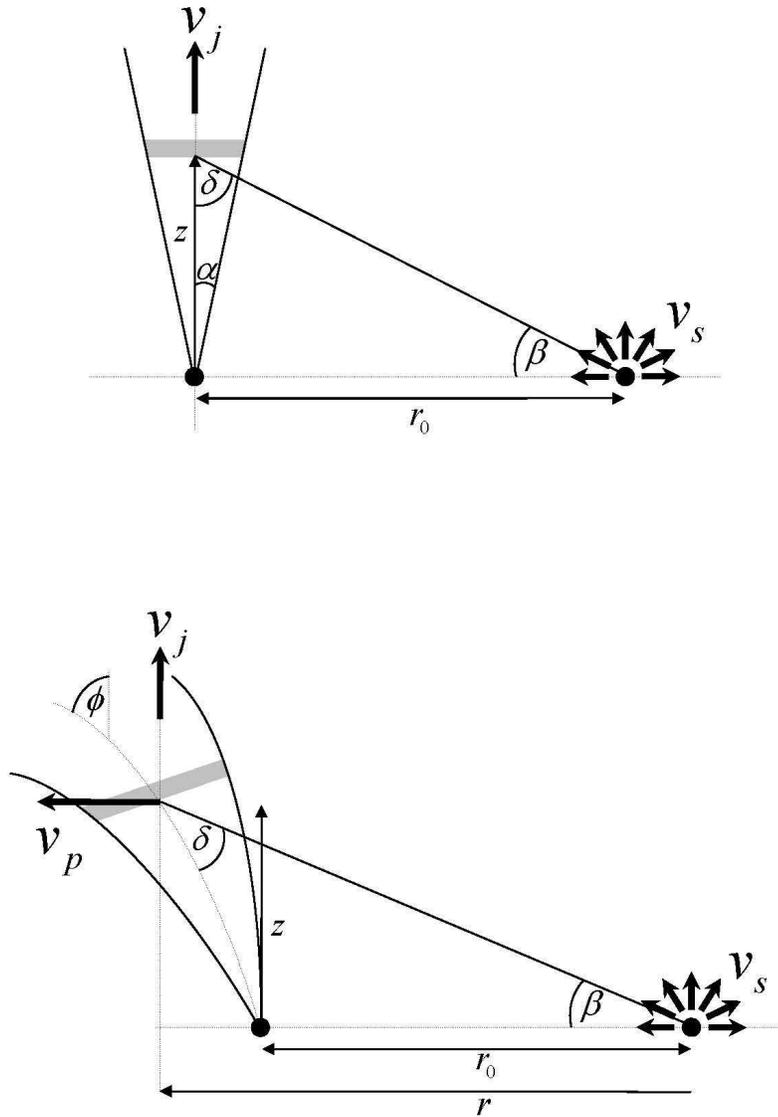} \vskip -3.2 cm
\caption{Schematic drawing of the interaction between the jet and the slow
wind that bends it. The slow wind is blown at velocity $v_s$ by the primary star
on the right, while the jet is blown by the accreting secondary star
on the left. The jet's speed $v_j$ is assumed to be $v_j \gg v_s$.
The mass loss rate in the two jets combined is $\dot M_j$, and the
mass loss rate of the slow wind is $\dot M_s$.
The orbital motion is ignored.
The upper panel shows the jet when no bending occurs for clear definition
of some quantities, while the lower panel show the bent jet.
An opposite jet exists on the other side of the equatorial plane. }
\label{draw1}
\end{figure}

The bending interaction is drawn schematically by Soker \& Rappaport (2000) and
Livio \& Soker (2001), and it is shown in Figure \ref{draw1};
3D images of numerical simulations are presented by Garcia-Arredondo \& Frank (2004).
The slow wind has a spherical mass loss rate of $\dot M_s$ and a relative
speed to the primary star of $v_s$.
A small fraction of this wind is accreted by the secondary star, forms an accretion
disk that blows two jets, with a mass loss rate of $\dot M_j$ into the
two jets together, and with a speed of $v_j$ perpendicular to the equatorial plane
relative to the secondary star.
Although the jets can have a large opening angle and, in many cases, are likely to
have a large opening angle, in the present study we assume a narrow jet with
a half opening angle $\alpha \ll 1$, and also assume that the jet is bent
as one entity (sound crossing time across the jet is very short).
The density per unit length along the jet axis is
\begin{equation}
m_j = \frac {\dot M_j}{2 v_j}
\label{mj}
\end{equation}
(recall that $\dot M_j$ is the mass loss rate into the two jets together).

We move to a frame of reference attached to the secondary star in its orbital
motion, with a velocity relative to the primary of
${\bf {v}}_{\rm orb}=v_r \hat r + v_\theta \hat \theta$, where
$v_\theta \simeq r \dot \theta$, $\theta$ is the relative angle of the two stars
in the equatorial plane, and $r$ is the projected distance from the primary to the jet
on the equatorial plane; $v_r<0$ when the two stars approach each other.
We consider a narrow  jet's segment at a height $z$ above (or below) the equatorial plane.
The slow wind segment that hit this segment left the primary at an angle $\beta$
to the equatorial plane (see fig. \ref{draw1})
\begin{equation}
\sin \beta = \frac{z} {(z^2+r^2)^{1/2}}.
\label{stheta}
\end{equation}
The slow wind that hits the jet at a high $z$ above the equatorial plane has
a relative velocity to the jet of
\begin{equation}
v_{rel}= [v_\theta^2+(v_s \cos \beta-v_r)^2+(v_s \sin \beta)^2]^{1/2}.
\label{vrel}
\end{equation}
We consider a fast jet $v_j \gg v_s$ that initially expands perpendicularly
to the orbital plane, but is then bent by the ram pressure of the slow wind
and acquires a velocity parallel to the equatorial plane $v_p$.
The ram pressure exerted by the slow wind on the jet in a direction
parallel to the equatorial plane is
\begin{equation}
P_{ram}=
\rho \left\{ [v_\theta^2+(v_s \cos \beta-v_r)^2]^{1/2}-v_p \right\}^2,
\label{pram1}
\end{equation}
where the density of the slow wind
\begin{equation}
\rho = \frac {\dot M_s}{4 \pi v_s (r^2+z^2)}.
\label{rho}
\end{equation}

The equation for accelerating the jet in a direction parallel to the equatorial
plane (perpendicular to the initial direction of the jet), under the assumption
of a fast jet, $v_p \ll v_j$, reads
\begin{equation}
\frac{dv_p}{dt}= \frac{P_{ram} 2 z \tan \alpha}{m_j}
\label{dvpdt}
\end{equation}
Under the assumption of a fast jet, $z=v_j t$ and $dt=dz/v_j$.
We also scale velocities by the slow wind speed
\begin{equation}
u_r \equiv \frac{v_r}{v_s}; \quad u_\theta \equiv \frac{v_\theta}{v_s}; \quad
u_p \equiv \frac{v_p}{v_s}; \quad u_j \equiv \frac{v_j}{v_s}.
\label{defvs}
\end{equation}
The equation of motion reads
\begin{equation}
\frac{du_p}{dz}= A \left\{
\left[ u_\theta^2+\left( \frac{r}{\sqrt{r^2+z^2}}-u_r \right)^2 \right]^{1/2}
-u_p \right\}^2 \frac{z}{r^2+z^2},
\label{dvpdz}
\end{equation}
where
\begin{equation}
A= \frac{\tan \alpha}{\pi} \frac{\dot M_s}{\dot M_j}
\label{adef}
\end{equation}

The meaning of the different terms in equation (\ref{dvpdz}) are as follows.
(1) The factor $A$ is proportional to the ratio of colliding masses.
Bending efficiency increases with $A$.
(2) The terms $u_\theta$ and $u_r$ result from the orbital motion of the
secondary star, which blows the jets, relative to the slow wind.
(3) The term $r/(r^2+z^2)^{1/2}$ results from the ram pressure of the
slow wind on the jet. The slow wind moves at a velocity $v_s$; but since velocity was
scaled by $v_s$, a factor of unity multiplies this term.
(4) The numerator in the last term is due to the increase in the jet cross section,
and it increases the bending efficiency as the jet expands.
(5) The denominator in the last term is the decrease in the slow wind density,
and it makes bending less efficient as distance from the primary star grows.

\subsection{Results for Impulsive Jets}

We consider a case in which the jets are blown by a secondary star
that is less massive than the primary star.
companion. The slow wind is blown by the primary star, residing close to the
center of mass of the binary system.
The formulation derived above is applicable to continuously blown jets, or
jets blown impulsively.
However, for the bubbles in PNs, or similar object, to be significantly
displaced by the mechanism discussed in $\S 2.1$, the jet should be
blown during a short time compare to the orbital period.
(Significant displacement from axisymmetry for continuously blown jets can
be acquired if the binary system has a large eccentricity;
see references in Soker \& Hadar 2002.).
In many PNs, the jets' ejection (PN jets refer to the jets blown
by the PN progenitor) can take place over a short time period
(e.g., Meaburn 2006), which we take to be shorter than the orbital period.
For example, the orbital period can be 5-50 years (orbital separation of
$\sim 3-20 \AU$), and the ejection event a few years, as in
symbiotic-nova outbursts on an accreting WD companion.
The mass accretion rate from the primary stellar wind, $\dot M_2$, by a companion
of mass $M_2$ at an orbital separation $r_0$ is
\begin{equation}
\frac{\dot M_2}{\dot M_s} \simeq
0.05
\left( \frac {M_2}{0.6 M_\odot} \right)^{2}
\left( \frac {v_{rel}}{15 \km \s^{-1}} \right)^{-4}
\left( \frac {r_0}{10 \AU} \right)^{-2}.
\end{equation}
If in impulsive jets' ejection $\dot M_j \sim 0.2 \dot M_2$, then
for the above mass accretion rate $A \simeq  5.6 \alpha/10^\circ$.
In short eruption events, like disk instability or nova-like outbursts on an
accreting WD, it might be that $\dot M_j > 0.2 \dot M_2$, and $r_0$ span a
range of $\sim 1-30 \AU$.
Therefore, we consider $A$ to be be in the range $A \sim 0.1-100$.

The jet is bent, according to equation (\ref{dvpdz}), and $u_p$, the velocity component
parallel to the equatorial plane and perpendicular to initial velocity of the jet
reaches an asymptotic velocity of
\begin{equation}
u_{pa} = \left[ u_\theta^2+\left( \frac{r}{\sqrt{r^2+z^2}}-u_r \right)^2 \right]^{1/2}.
\label{upa}
\end{equation}
where $u_{pa}$ is in unit of the slow wind speed $v_s$.
The asymptotic (final) velocity $u_p$ due to the orbital tangential velocity $u_\theta$
does not depend on the factor $A$ or the jet speed $v_j$ (or $u_j=v_j/v_s$).
This is approximately true for the radial orbital component $u_r$ as well,
meaning that the initial jet velocity component along the secondary stellar orbital
motion is quite efficiently reduced to zero.
The departure from axisymmetry due to the orbital motion of the star blowing
the jet will be small.
Therefore, in imposing a noticeable large-scale departure from axisymmetry,
where the two jets are bent to the same side, the bending due to the
slow wind outflow from the primary star must be considered.
This bending is less efficient because the slow wind velocity is
not perpendicular to the jet velocity after the jet leaves the equatorial
plane, as seen by the decreasing of the term $r/(r^2+z^2)^{1/2}$.
Ignoring the orbital motion, equation (\ref{dvpdz}) reads
\begin{equation}
\frac{du_p}{dz}= A
\left[  \frac{r}{(r^2+z^2)^{1/2}} -u_p \right]^2  \frac{z}{r^2+z^2},
\label{dup}
\end{equation}
This equation is supplemented by another equation for the jet propagation
along the direction perpendicular to the equatorial plane.
For a fast jet, $v_j \gg v_s$, this reads,
\begin{equation}
\frac{dr}{dz}= \frac{u_p}{u_j}.
\label{dr}
\end{equation}

\begin{figure}
\vskip -3.2 cm
\hskip -1.30 cm
\includegraphics[width =170mm]{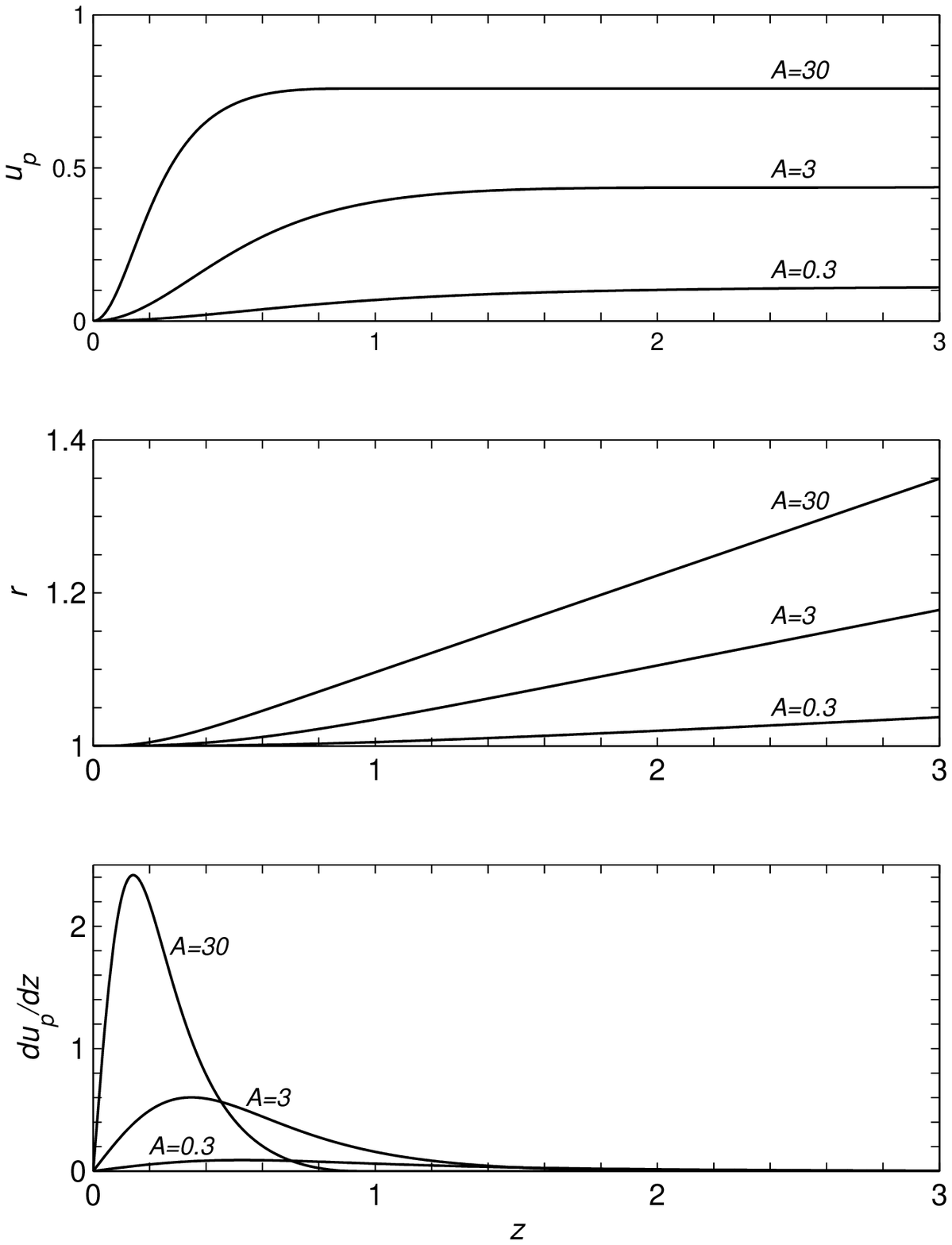} \vskip -3.2 cm
\caption{The bending properties of a jet with initial velocity,
$u_j \equiv v_j/v_s=6$, and for three values of $A$ as defined in
equation (\ref {adef}). The distance $z$ form the equatorial plane
is given in units of the binary orbital separation $r_0$.
{\it Upper panel:} The transverse (to the initial jet velocity) jets' velocity,
in units of the slow wind velocity ($u_p \equiv v_p/v_s$), obtained by solving
the coupled equations (\ref{dup}) and (\ref{dr}).
{\it Middle panel:} The projected distance on the orbital plane $r$ in
units of $r_0$ (see fig. \ref{draw1}).
{\it Lower panel:} The transverse (bending) acceleration ${du_p}/{dz}$
as given in equation (\ref{dup}).}
\label{upf1}
\end{figure}

\begin{figure}
\vskip -4.2 cm
\includegraphics[width =180mm]{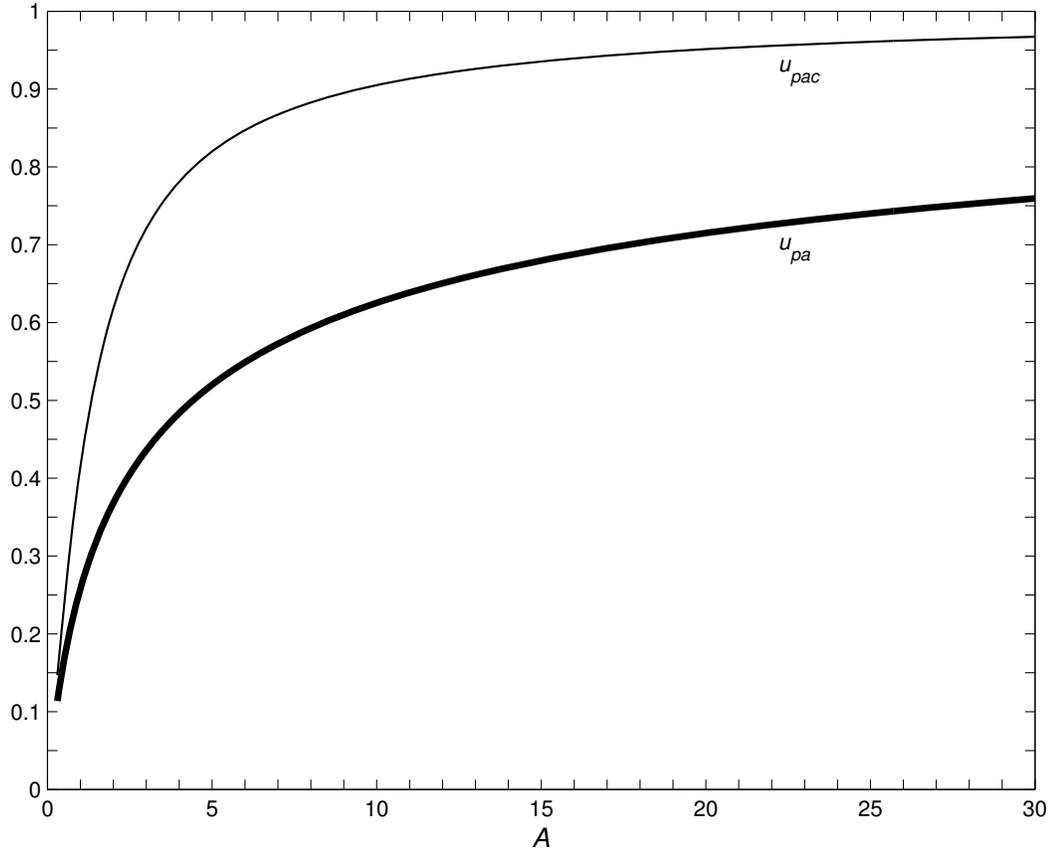} \vskip -3.2 cm
\caption{The asymptotic transverse velocity of $u_p$, $u_{pa}$,
as function of the factor $A$ (eq. \ref{adef}) is depicted by the
thick line.
The thin line is the crude approximate equation for the asymptotic
transverse velocity as given by equation (\ref{upas}). }
\label{upaf}
\end{figure}

Figure \ref{upf1} presents the numerical solutions of the last two coupled equations
for initial jet's speed $u_j\equiv v_j/v_s=6$ and for three values of $A$
as function of the distance from the equatorial
plane $z$ in units of the orbital separation $r_0$.
The velocity $u_p$ is plotted in the upper panel, in the middle panel
the projection of the jet distance on the equatorial plane $r$ is drawn
(in  units of $r_0$), while the lower panel presents the acceleration $du_p/dz$.
In Figure \ref{upaf}, we show the asymptotic velocity $u_{pa}$ as a function of $A$
for $u_j=6$ (the thick line).

Changing the initial jet's speed $u_j$ does not change the solution
for $u_p$, while the quantity $r-r_0$ is proportional to $u_j^{-1}$, because the
bending angle is given by $\tan \phi =u_p/u_j$, so that for faster jets
the bending angle decreases; the dependence is $\phi \propto v_j^{-1}$.
This can be understood as follows. As the jet speed $v_j$ increases,
the time of accelerating the jet by the slow wind's ram pressure along
a distance $dz$ decreases as $v_j^{-1}$;
however, the density in the jet decreases as $v_j^{-1}$ as well.
Hence the total change in $v_p$ (or $u_p$) along a distance $dz$ does
not depend on $v_j$ under our assumptions, in particular the assumption
$v_j$ (or $u_j$) is constant and is not influenced by the
interaction with the slow wind.
As seen from Figure \ref{upf1}, for a fast jet (and for typical values used here)
$u_p$ almost reaches its terminal speed while $r \simeq r_0$, where $r_0$
is the initial orbital separation.
Note that most of the bending occurs for $r$ not much larger than $r_0$.
For $r=r_0$ equation (\ref{dup}) reads then
\begin{equation}
\frac{du_p}{dz}= A
\left[  \frac{r_0}{(r_0^2+z^2)^{1/2}} -u_p \right]^2  \frac{z}{r_0^2+z^2},
\label{dup0}
\end{equation}
The solution near the origin, when $u_p \ll 1$ is
\begin{equation}
u_{p0} (r \sim r_0) \simeq
 \frac{A}{2} \frac{z^2}{r_0^2+z^2}.
\label{up0}
\end{equation}
The asymptotic velocity is reached when the numerical value inside the
square brackets in equation (\ref{dup0}) is very small, or
\begin{equation}
u_{pac} (r \sim r_0) \simeq  \frac{r_0}{(r_0^2+z^2)^{1/2}}.
\label{upaa}
\end{equation}
The change of behavior  between the solution near the jets' origin
and the asymptotic solution takes place when $u_{p0} \sim u_{pa}$, which
by equations (\ref{up0}) and (\ref{upaa}) is
\begin{equation}
u_{pac} \sim   \left( 1+ \frac{1}{A^2} \right)^{1/2} -\frac{1}{A}.
\label{upas}
\end{equation}
This very crude expression for the asymptotic transverse velocity
is drawn by a thin line on Figure \ref{upaf}.

As the jet leaves the launching accretion disk, it is very dense and no
bending occurs, namely, $du_p/dz$ is very small.
At large distances from the jet's origin, the angle $\delta$ is small and bending
is no longer efficient.
The bending is most efficient at some intermediate value of $z$, after the density of
the jets decreases as they expand, but before the angle $\delta$ decreases much.
Practically, this intermediate value of $z$ is quite close to the jet's origin,
$z \la r_0$, as is seen in the lower panel of Figure \ref{upf1}.

If the jet pair in a binary system is known to be blown in a time period much shorter
than the orbital period (an impulsive jet pair),
and a bending is observed, the bending angle can be used with the thick line
in Figure \ref{upaf} to find the constant $A$ given in equation (\ref{adef}).
Thus the relation between the three quantities:
the primary stellar mass loss rate $\dot M_s$, the secondary stellar
mass loss rate to the two jet $\dot M_j$, and the half opening angle of
the jets $\alpha$, can be found.

\section{BENDING IN CLUSTERS}
\label{clben}

Many jet pairs blown by radio galaxies are observed to be bent as a result of
the relative motion of the galaxy and the ICM (e.g., Bliton et al. 1998).
Radio jets which are strongly bent are called narrow-angle tailed (NAT)
radio galaxies, while those with slightly bent jets are called
wide-angle tailed (WAT) radio galaxies.
Many of the WAT radio galaxies are dominant galaxies in clusters, like
cD galaxies (Owen \& Rudnick 1976; Burns et al. 1979).
A bulk motion of the IC, e.g., as a result of cluster merger, can efficiently
bend radio jets (Bliton et al. 1998).
A bulk ICM motion relative to the central cD galaxy can exist as a result of merging
with a sub-cluster (group of galaxies), as found in several cases
(Dupke \& Bregman 2005, 2006; Fujita et al. 2006).

The bending process of jets by the ICM was extensively studied
(e.g. Balsara  \& Norman  1992); the calculations are not repeated here.
Basically, the bending of jets in clusters is characterized by the
curvature radius $R_c$ of the bent jet.
Because the ambient density changes slowly with distance from the cluster center,
unlike the case in PNs, a constant ambient density is assumed in the region
where most of the jet's bending occurs.
An approximate expression for the radius of curvature is (Sarazin et al. 1995)
\begin{equation}
R_{\rm {curv}} \sim 2
\left( \frac {L_{2j}}{3 \times 10^{43} \erg \s^{-1}} \right)
\left( \frac {R_j}{1 \kpc} \right)^{-1}
\left( \frac {v_j}{0.1 c} \right)^{-1}
\left( \frac {v_a}{300 \km \s^{-1}} \right)^{-2}
\left( \frac {n_e}{0.1 \cm^{-3}} \right)^{-1} \kpc
\label{rcur}
\end{equation}
where $n_e$ is the ambient electron density,
$R_j$ is the jet's radius where most bending takes place, the total
mechanical (kinetic) power of the two jets, $L_{2j}$, was scaled according to
Birzan et al. (2004), and the relative ambient to jet speed $v_a$ according to
Malumuth (1992), with a $90^\circ$ angle between the
initial jet velocity and the ambient flow.
To achieve a noticeable bending in cD clusters we require $R_{\rm curv} \la 10 \kpc$.
This implies that even if the relative velocity of the cD galaxy to the
ambient medium component perpendicular to the jet axis is $\sim 100 \km \s^{-1}$
we get the required bending, as observed in some jets, or bubbles, blown by cD
galaxies.
This required bending will occur also for a narrower jet or a faster jet with
$v_j$ close to the speed of light $c$,

Our primary interest is to compare the bending process in binary stars to that of
jets blown by the dominant cD galaxies at the centers of cooling flow clusters
(or galaxies).
Common to bending of jets in binary progenitors of PNs (referred to as bending
in PNs) and bending of jets at the centers of clusters of galaxies
is that the jets are bent by the ram pressure due to the relative motion of
the medium the jets expand to and interact with, both in PNs and clusters.
This is unlike point-symmetric structures which result from precession; in
both PNs and clusters precession is due to the accretion disk that launches
the jets and not due to the ambient medium.
There are some differences between bending in clusters and bending in binary systems,
such as PNs' progenitors.
\begin{enumerate}
\item {\it Relative velocity.} The bending considered in PNs is due to the
outflow velocity of the slow wind blown by the primary star. This implies that the
angle between the jets velocity and the ambient medium velocity, $\delta$ in
Figure \ref{draw1}, decreases very fast. It decreases even if
bending does not occur.
In clusters the velocity is due to the motion of the cD galaxy relative to the ICM.
The angle decreases only because of the bending, and it decreases slowly.
\item {\it Densities.} The jet's density decreases as the jet moves outward,
both in clusters and in stellar binary systems.
However, in PNs the density of the bending slow wind
(see Fig. \ref{draw1}) decreases as well, $\rho_s \propto (r^2+z^2)$
(denominator of last term in eq. \ref{dvpdz}), while near the center of clusters the
ICM density profile is much shallower and decreases slowly with increasing distance.
\item {\it Asymmetry in clusters.} The bending process in PNs is the same for the
two opposite jets, but this is not necessarily the case in clusters of galaxies.
If the jets are not blown perpendicular to the relative velocity between the ICM
and the galaxy, then the jet expanding to the same direction the galaxy moves
to will feel a larger ram pressure opposing its expansion velocity,
and it will be slowed down more efficiently.
More important, as this jet is bent, the angle of the relative velocity between
the ICM and the galaxy to the jet's axis will increase to $90^\circ$
before decreasing.
In the opposite jet
this angle decreases continuously. Therefore, the jet expanding against the ICM motion
will be bent more than the other jet.
Asymmetry between the two jets in clusters can be also caused by the presence of
asymmetric strong magnetic fields in the ICM (Soker 1997),
and/or density inhomogeneities such as clouds (Sarazin et al. 1995).
\item {\it Late stages of the bending process. }
After reaching their asymptotic bending angle $\phi$, jets in binary stellar systems
will not bend any more.
If a jet inflates a bubble, it will move outward radially along the streaming
slow wind material.
In clusters the situation is different because of the flow structure mentioned in
points (1) and (2) above and because the low density bubbles buoyant outward.
The result is that although the radio jets of cD galaxies are not bent much,
after they become subsonic the bending is very efficient
(Eilek et al. 1984; Odea \& Owen 1986), and the asymmetry between the two sides
can substantially increase (Burns et al. 1986).
\item{\it The effective bending location.} From the differences in points 1,2, and 4
it turns out that in stellar binary systems most of the bending occurs close to the
jets' origin (lower panel of fig. \ref{upf1}).
In clusters the bending becomes more efficient as the jet expands and slows down.
In particular, if the jet inflate bubbles, they move slowly, have very low density,
and large cross section. Thus, in clusters the departure from axisymmetry will be most noticeable
in bubbles.
\end{enumerate}

Despite the differences listed above, there are some striking morphological
similarities of bubbles displaced from the symmetry axis in PNs and clusters;
two cases are mentioned in $\S 1$ (see appendix of the astro-ph version of the
paper).

\section{SUMMARY}

In recent years the jet shaping model for many PNs and similar objects,
like the massive binary star $\eta$ Carinae, acquired considerable acceptance.
It should be stressed that not all PNs were shaped by jets, and not
all morphological structure in PNs were formed by jets.
Bubble pairs, though, are most likely inflated by double jets, and
the jets are probably blown by a stellar secondary star.
The secondary star accretes mass from the primary's slow wind, forms an accretion
disk and blow two jets, either continuously or impulsively,
that is, during a time shorter than the orbital period.
In some PNs, the line joining the centers of the two bubbles in a pair
does not pass through the center of the nebula, meaning that the bubbles are
displaced such that the nebular structure departs from axisymmetry.
The explanation is that the two jets that inflate the bubbles were bent to
the same side by the ram pressure of the slow wind (Fig. \ref{draw1}).

We therefore set the goal of deriving a simple and approximate relation between
the bending angle of the jets and the properties of secondary stellar jets and the
primary slow wind.
For fast jets, $v_j \gg v_s$, the important factor is the quantity $A$
defined in equation (\ref{adef}).
The relation between the jet's asymptotic transverse speed $v_{pa}$
(see fig. \ref{draw1}) and $A$ is presented by the thick line in
Figure \ref{upaf}, and a very crude approximation is given in equation (\ref{upas})
($v_{pa}$ is in units of the slow wind speed $v_s$).
If the jets are impulsive, then the bending will be easier to observe;
otherwise it is averaged over different directions as the binary system rotate.
If $A$ is not too small, and the orientation of the nebula is such that the
bending is not along the line of sight, then observations may reveal
the two jets or the bubbles (lobes) inflated by the jets to be displaced to
the same side of the symmetry axis.
Examples of such PNs are listed and classified by Soker \& Hadar (2002).

In some clusters, X-ray-deficient bubble (cavity) pairs that were inflated by
jets blown by the central cD galaxy, show displacement from axisymmetry similar
to visible-deficient bubble (lobe) pairs observed in PNs (see appendix
of the astro-ph version of this paper).
We therefore set a second goal of comparing the bending process of jets
in these two groups of objects.
Two factors of the bending process are common to these two classes of objects:
1) the bending results from the ram pressure perpendicular to the jet axis,
and 2) the ram pressure is exerted by the same external medium the jets
expand to and interact with.

However, there are some significant differences listed in \S\ref{clben}.
(1) Because the bending in binary systems results from the slow wind blown by the
primary star, the ambient density decreases faster with distance than the ambient
density of the ICM in the centers of clusters.
(2) Also, the angle $\delta$ (see Fig. \ref{draw1}) between the jet velocity
and ambient slow wind velocity in binary systems decreases with distance
along the jet axis, even when bending does not occur.
In clusters the relative velocity is due to the bulk ICM motion and
changes only because of bending.
(3) In binary systems the two opposite jets are likely to be blown perpendicular
to the orbital plane, thus they will be bent in the same way.
In clusters, the jets' axis need not be perpendicular relative to the bulk
motion of the ICM relative to the central black hole that blows the jets, so
the jet facing the ICM flow will be bent more efficiently.
In binary systems such an asymmetry between the two jets can occur if the
jets (more specifically the accretion disk that launches the jets) precess.
(4) In clusters,  after the bubbles (cavities; lobes) are inflated, they buoy outward.
They are more susceptible than the jets to the ram pressure, and departure from
axisymmetry may substantially increase. This process does not exist in binary systems
because the circumbinary ambient matter is not in hydrostatic equilibrium, but rather
the ambient matter expands at a high Mach number.
(5) In binary stars most of the bending occurs when the jets are at a distance
$z \la r_0$, where $r_0$ is the orbital separation.
In clusters the bending becomes more efficient at larger and larger distances.

We hope that the study presented in this paper will motivate researchers to
pay more attention to the departure from axisymmetry of bubble (cavity; lobe)
pairs in both clusters of galaxies and PNs.

\acknowledgments

\acknowledgments This research was supported in part by the Asher
Fund for Space Research at the Technion.



\newpage

\centerline{\bf APPENDIX}

In this {\it appendix} we summarize the morphological similarities between X-ray
deficient bubbles (cavities) in clusters or groups of galaxies and
visible-light deficient bubbles (lobes) in PNs.
The Table below is based on that given in Paper-1, with some additional comparisons.
In Paper 1, Soker also pointed out similar values of some non-dimensional
quantities between clusters and PNs.
These similarities led Soker to postulate a similar formation mechanism,
thereby strengthening models for PN shaping by jets, although not all PNs
are shaped by jets; the jets in PN progenitors are likely blown by binary companions.

The images of the objects listed in Table 1 of Paper-1 are summarized in a
{\it PowerPoint} file Soker presented at the Asymmetrical Planetary Nebulae III
meeting (2003) at
\newline
http://www.astro.washington.edu/balick/APN/APN\_talks\_posters.html
\newline
go the `ppt' file in the ``Discussion'' of Session 13.
Other images were added here.
All image sources are listed, and references not in the list of the paper
are listed after the table.

\newpage
\begin{deluxetable}{lll}
\tablewidth{0pt}
\tablehead{
\colhead{Structure} & \colhead{Clusters} & \colhead{PNs}
}
\startdata 
Butterfly shape; faint  & Abell 478  [1]$^4$     & Roberts 22 [2]         \\
along symmetry axis$^1$ & (Sun et al. 2003) & (Sahai et al. 1999) \\
\hline  
Pairs of fat spherical & Perseus  [3]        & NGC 3587  [4] \\
bubbles near center    & (Fabian et al. 2000)  & (Guerrero et al. 2003) \\
\hline  
Closed bubbles  connected       & Abell 2052  [5]          &   VV 171  [6]       \\
at the equatorial plane      & (Blanton et al. 2001) & (Sahai 2001)   \\
\hline  
Open bubbles           & M 84                  &  He 2-104           \\
connected at the       & (Finoguenov \& Jones  & (Sahai \& Trauger,  \\
equatorial plane       & 2001) [7]       &  1998) [8]          \\
\hline 
Pair of bubbles detached   & HCG 62 [9]            &  Hu 2-1 [10]               \\
from a bright center      & (Vrtilek et al. 2002)    &  (Miranda et al. 2001b)  \\
\hline 
Point symmetry;     & MS 0735.6+7421 [11]           &  Hb 5 [12]$^3$ \\
suggesting precession$^{2}$ & (McNamara et al. 2005) &  (Terzian \& Hajian 2000) \\
\hline 
Bending to  &  HCG 62 [9]               & NGC 6886 [13]  \\
one side    & (Vrtilek et al. 2002)       & (Terzian \& Hajian 2000) \\
\hline 
point-symmetric     & Hydra A [14]                  & NGC 6537 [15]  \\
elongated lobes     & (McNamara et al.  2000)    & (Balick 2000)   \\
\hline 
Pairs of bright bullets     &  Cygnus A  [16]            &  NGC 7009  [17]     \\
along the symmetry axis   &  (Smith et al. 2002) &  (Balick et al. 1998)        \\
\hline 
Ripples             & Perseus [18]               &   M 57 (NGC 6720) [19] \\
                    & (Fabian et al. 2003;06) &   (Hora et al. 2005) \\
\enddata
\end{deluxetable}

Comments to the table in the {\it Appendix}:
\newline
{\bf (1)} Similar images of bubbles in clusters of galaxies and planetary nebulae (PNs).
In clusters these are X-ray images (e.g., with X-ray deficient bubbles),
while in PNs they are visible-light images (e.g., with visible-light deficient bubbles).
In the first eight pairs of images the similarity is of high degree.
In the last two pairs of images the similarity
between the cluster and the PN is of lesser degree.
\newline
{\bf(2)} See Pizzolato \& Soker (2005) for more detail.
\newline
{\bf(3)} The low-resolution image of the same object is from the
catalogue of Schwarz et al. (1992)
\newline
{\bf(4)} Free access to images are at these sites:
\newline
[1] http://arxiv.org/PS\_cache/astro-ph/pdf/0210/0210054.pdf
\newline
[2] http://ad.usno.navy.mil/pne/images/rob22.jpg
\newline
[3] http://arxiv.org/PS\_cache/astro-ph/pdf/0007/0007456.pdf
\newline
[4] http://arxiv.org/PS\_cache/astro-ph/pdf/0303/0303056.pdf
\newline
[5] http://arxiv.org/PS\_cache/astro-ph/pdf/0107/0107221.pdf
\newline
[6] http://ad.usno.navy.mil/pne/images/vv171.jpg
\newline
[7] http://arxiv.org/PS\_cache/astro-ph/pdf/0010/0010450.pdf
\newline
[8] http://ad.usno.navy.mil/pne/images/he2\_104.jpg
\newline
[9] http://chandra.harvard.edu/photo/cycle1/hcg62/index.html
\newline
[10a] http://arxiv.org/PS\_cache/astro-ph/pdf/0009/0009396.pdf
\newline
\indent see also: [10b] http://ad.usno.navy.mil/pne/images/hu21\_ha.gif
\newline
[11] http://arxiv.org/PS\_cache/astro-ph/pdf/0411/0411553.pdf
\newline
[12] http://ad.usno.navy.mil/pne/images/hb5.jpg
\newline
[13] http://ad.usno.navy.mil/pne/images/ngc6886.jpg
\newline
[14] http://arxiv.org/PS\_cache/astro-ph/pdf/0001/0001402.pdf
\newline
[15] http://ad.usno.navy.mil/pne/images/ngc6537.jpg
\newline
[16] http://www.journals.uchicago.edu/ApJ/journal/issues/ApJ/
\newline
\indent v565n1/54312/54312.web.pdf
\newline
[17a] http://ad.usno.navy.mil/pne/images/ngc7009.jpg
\newline
\indent see also (Goncalves et al.\ 2004, fig. 1):
\newline
[17b] http://arxiv.org/PS\_cache/astro-ph/pdf/0307/0307265.pdf
\newline
[18] http://chandra.harvard.edu/photo/2003/perseus/perseus\_ripple\_illustration\_nolabel.jpg
\newline
[19] http://www.spitzer.caltech.edu/Media/releases/ssc2005-07/ssc2005-07a.shtml

\begin{figure}
\includegraphics[width =150mm]{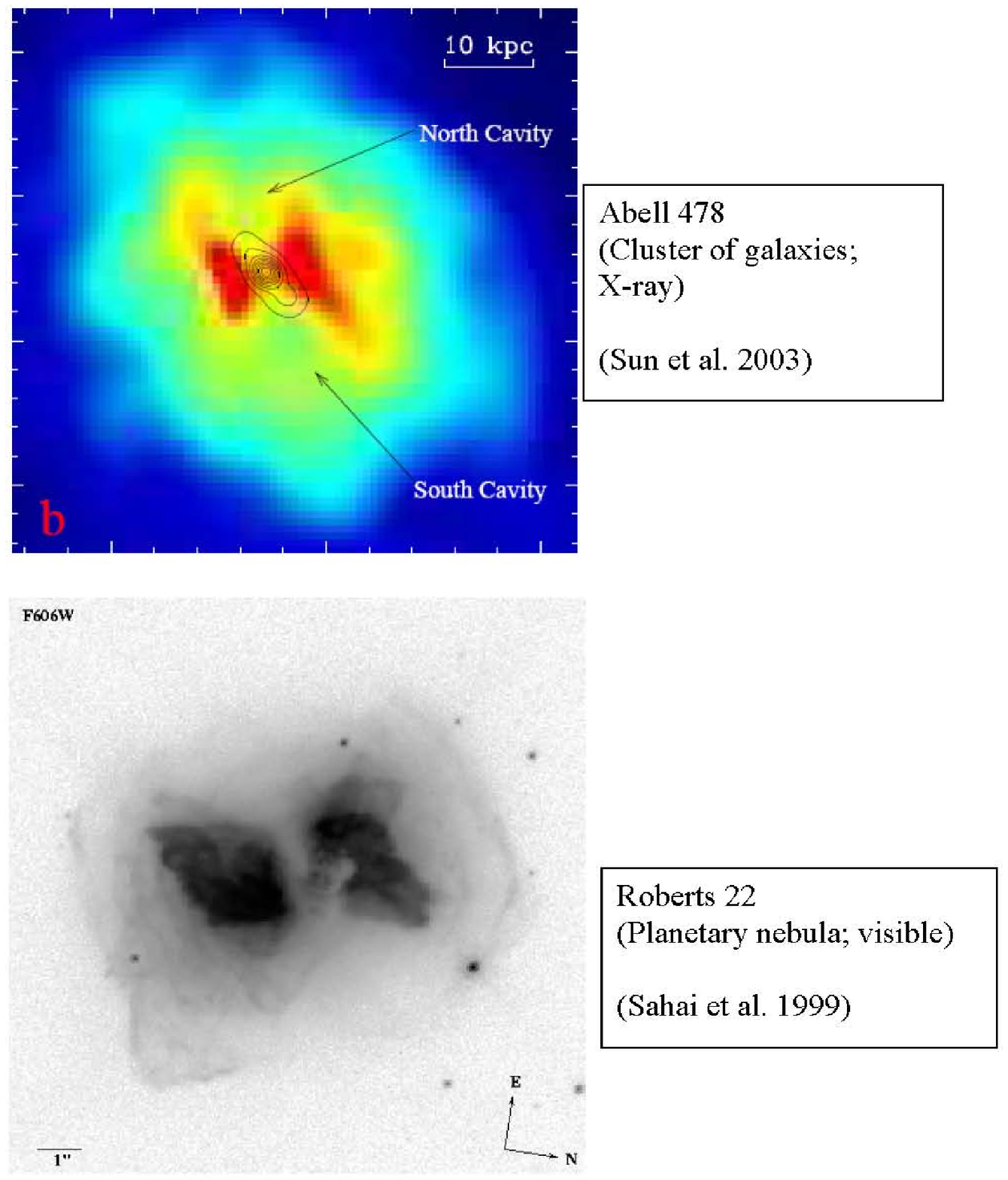}  \vskip -1.2 cm
\caption{Comparing false color X-ray image of a galaxy cluster with a
visible image of a planetary nebula, emphasizing the butterfly
shape and a faint region along the symmetry axis. }
\label{figap1}
\end{figure}
\begin{figure}
\includegraphics[width =150mm]{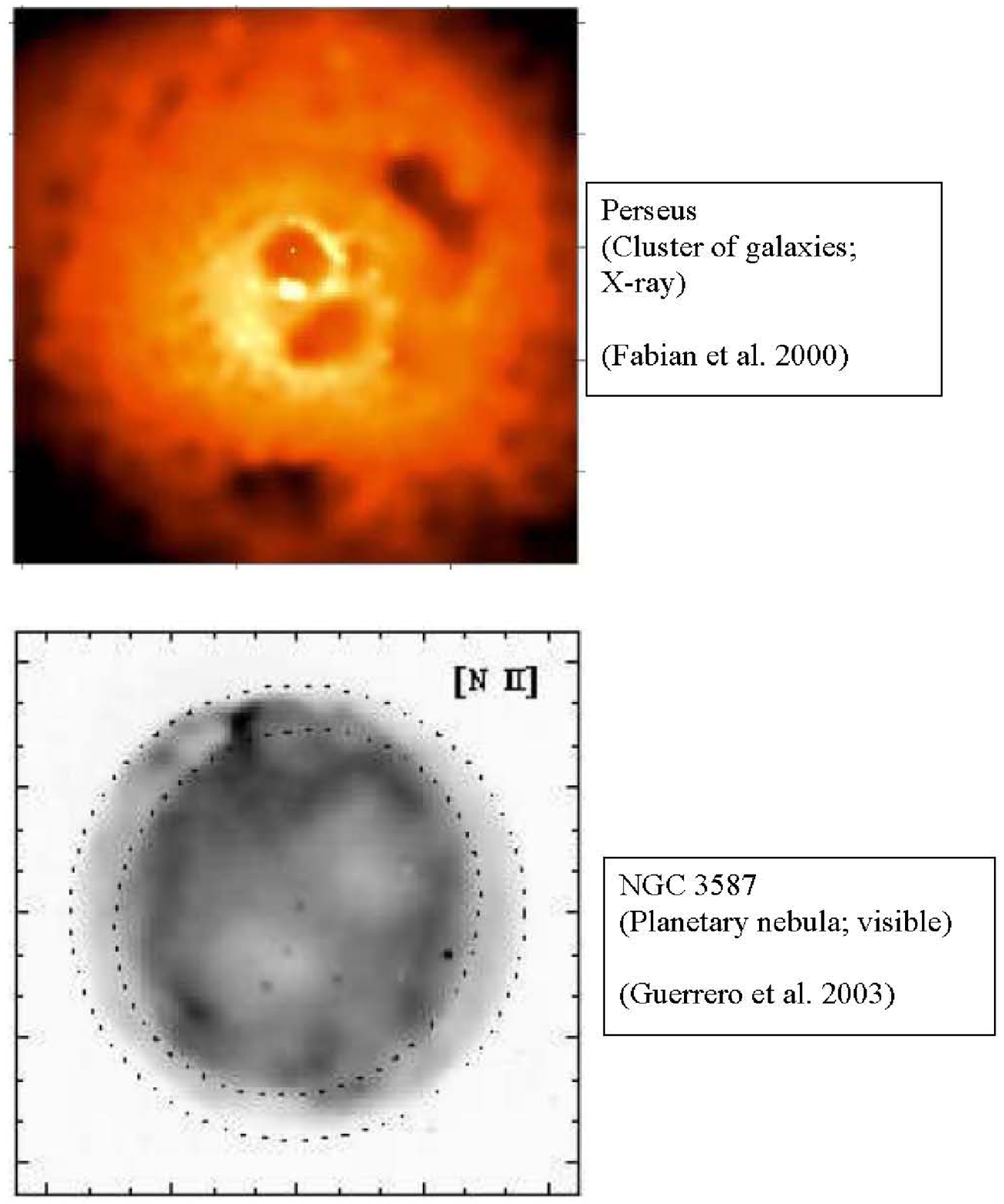} 
\caption{Like Figure \ref{figap1} but emphasizing
pairs of fat spherical bubbles near the center.}
\end{figure}
\begin{figure}
\includegraphics[width =150mm]{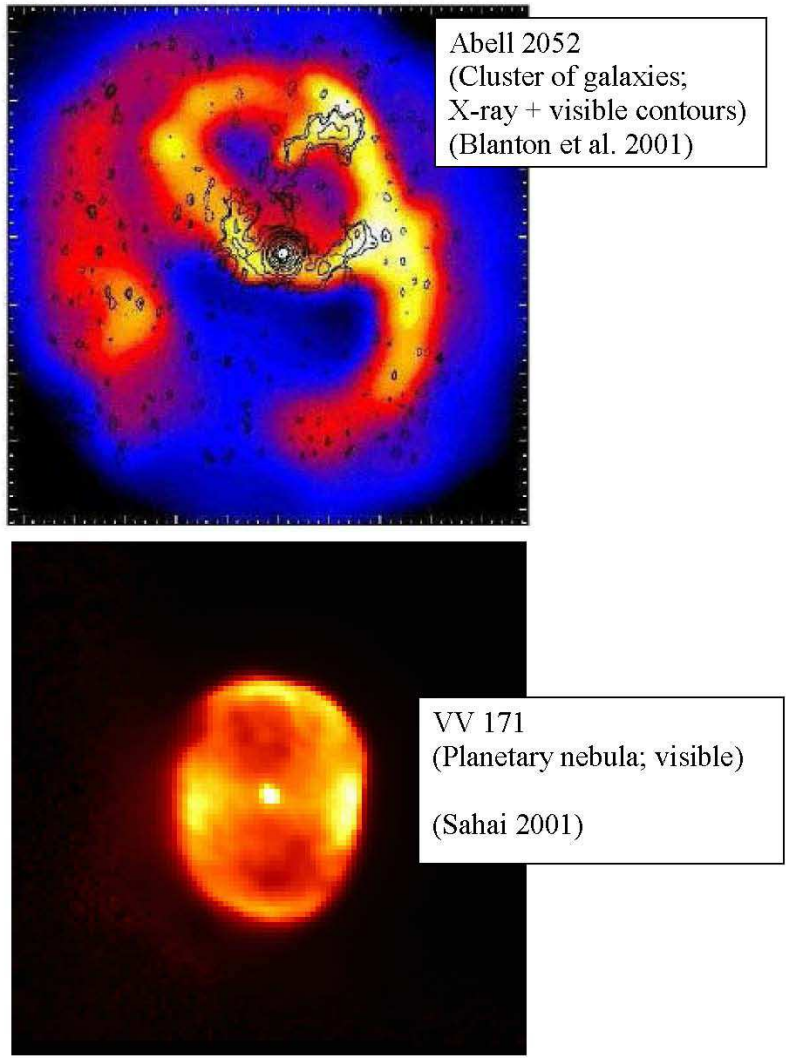} 
\caption{Like Figure \ref{figap1} but emphasizing
closed bubbles connected at the equatorial plane.}
\end{figure}
\begin{figure}
\includegraphics[width =150mm]{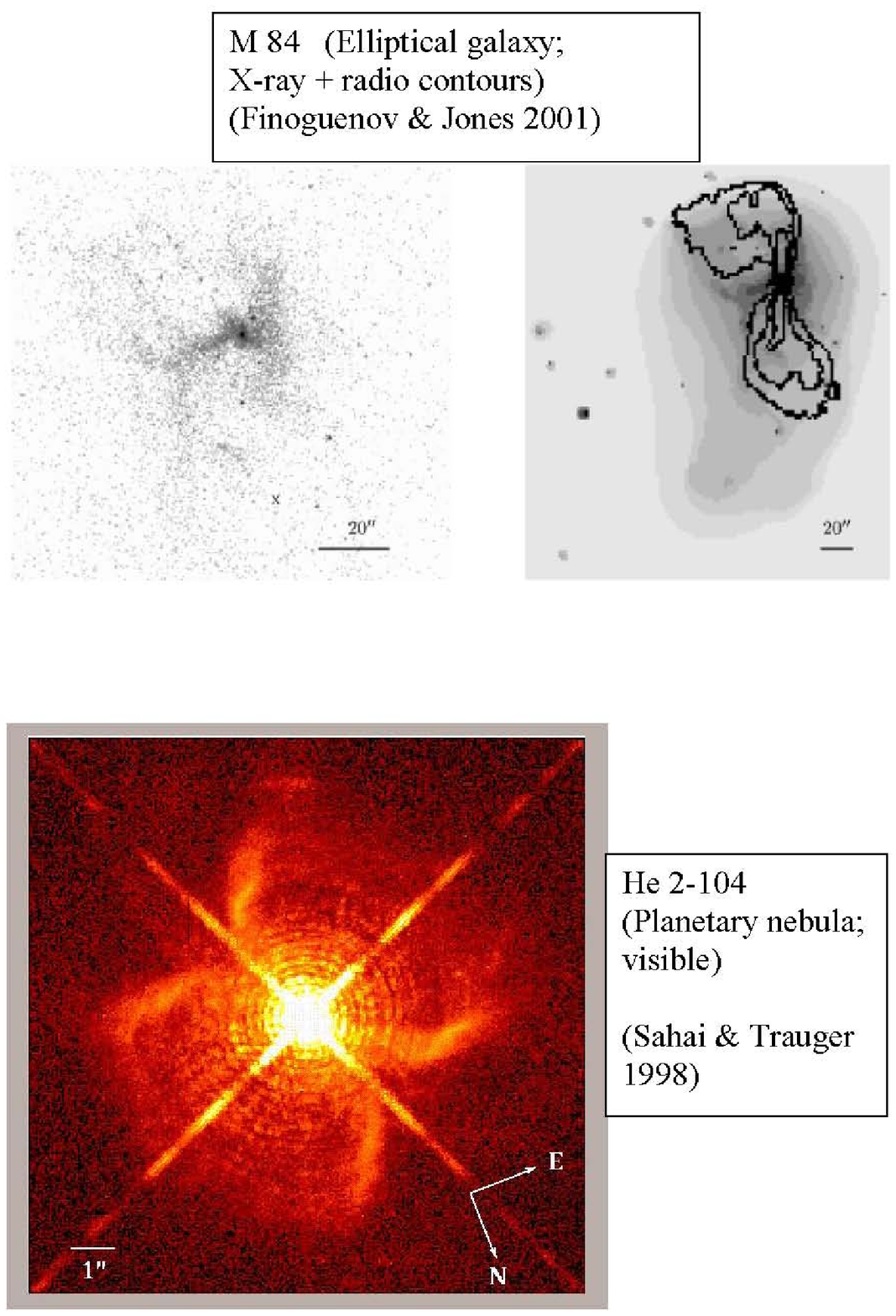} 
\caption{Like Figure \ref{figap1} but emphasizing
open bubbles connected at the equatorial plane.}
\end{figure}
\begin{figure}
\includegraphics[width =150mm]{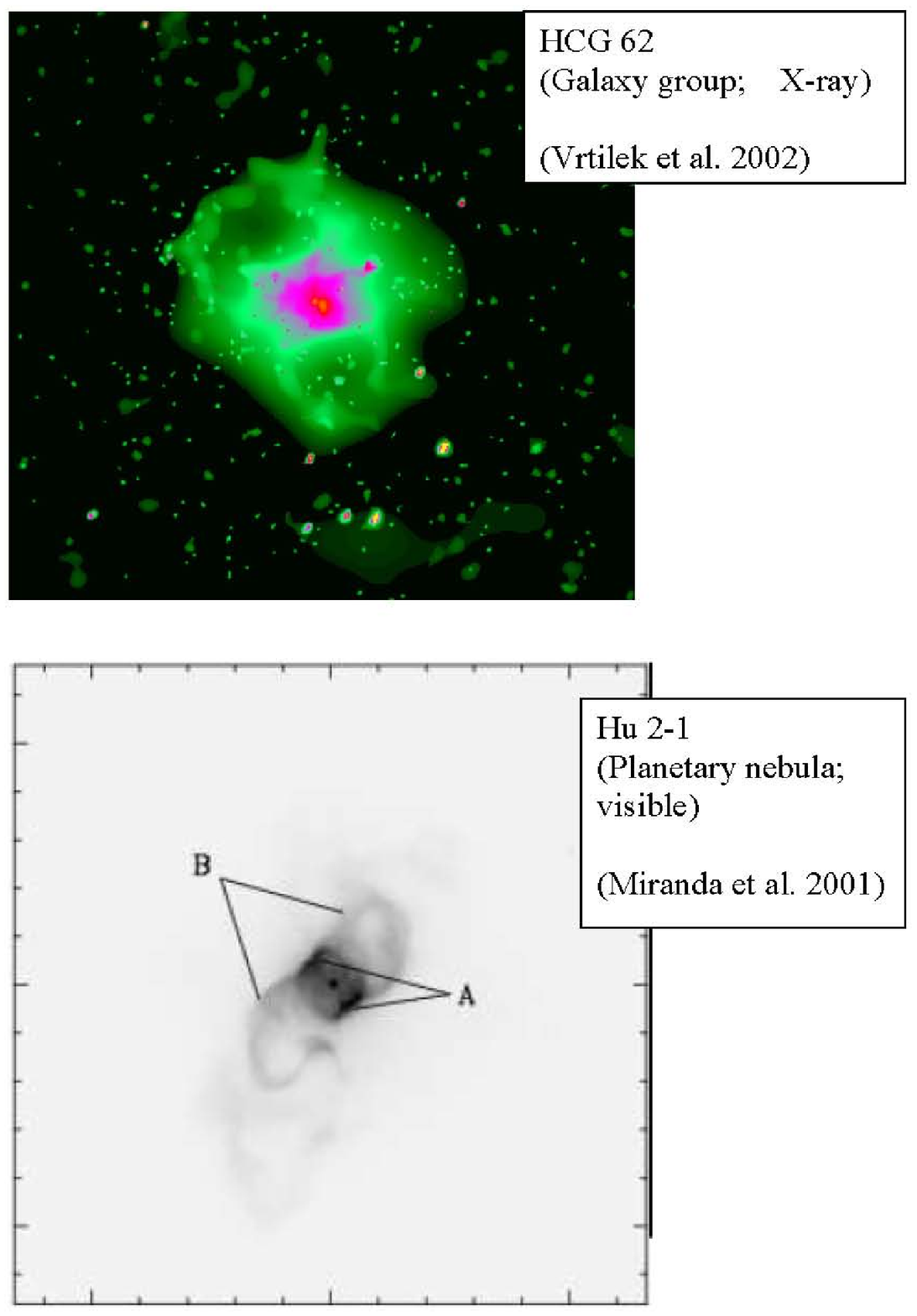} 
\caption{Like Figure \ref{figap1} but emphasizing a bubble pair
detached from a bright center .}
\end{figure}
\begin{figure}
\includegraphics[width =150mm]{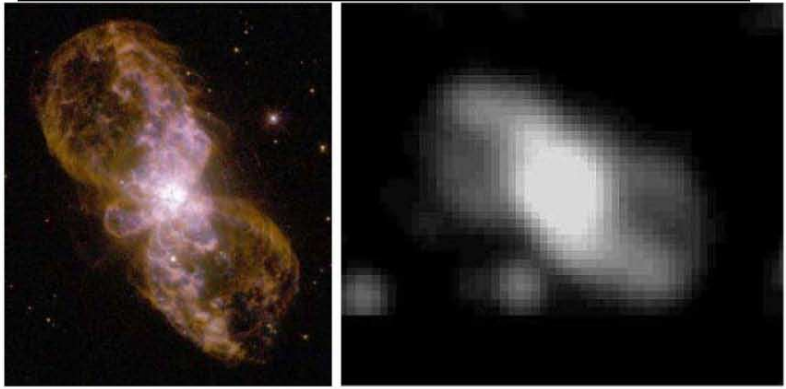} 
\caption{Like Figure \ref{figap1} but emphasizing
point symmetry.}
\end{figure}
\begin{figure}
\includegraphics[width =150mm]{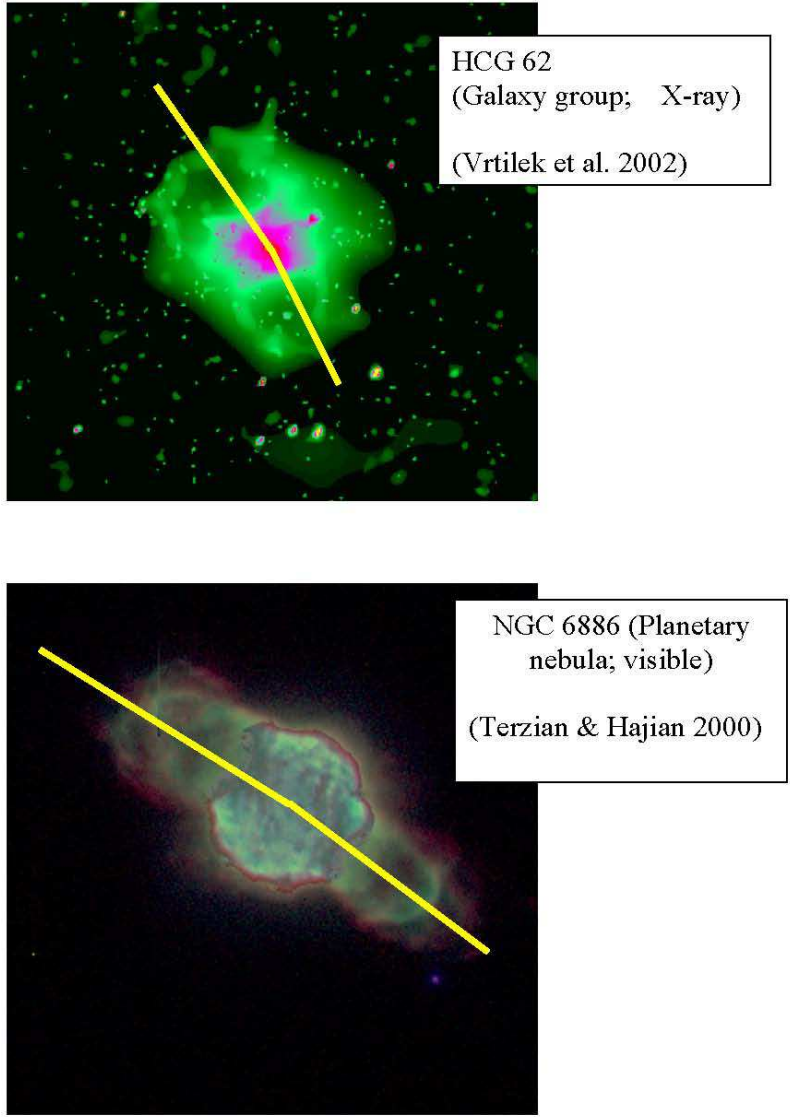} 
\caption{Like Figure \ref{figap1} but emphasizing
bending both bubbles (cavities; lobes) to one side (departure from axisymmetry).}
\end{figure}
\begin{figure}
\includegraphics[width =150mm]{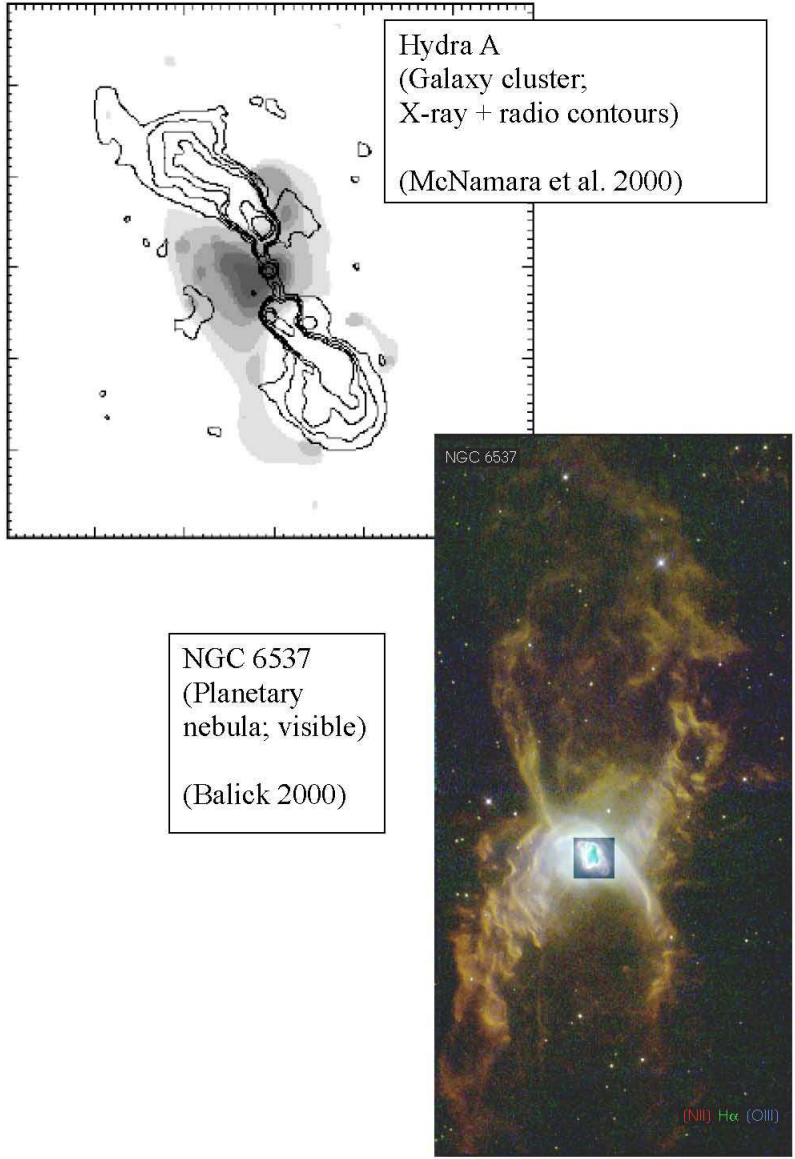} 
\caption{Like Figure \ref{figap1} but emphasizing
 point-symmetric elongated lobes.}
\end{figure}
\begin{figure}
\includegraphics[width =150mm]{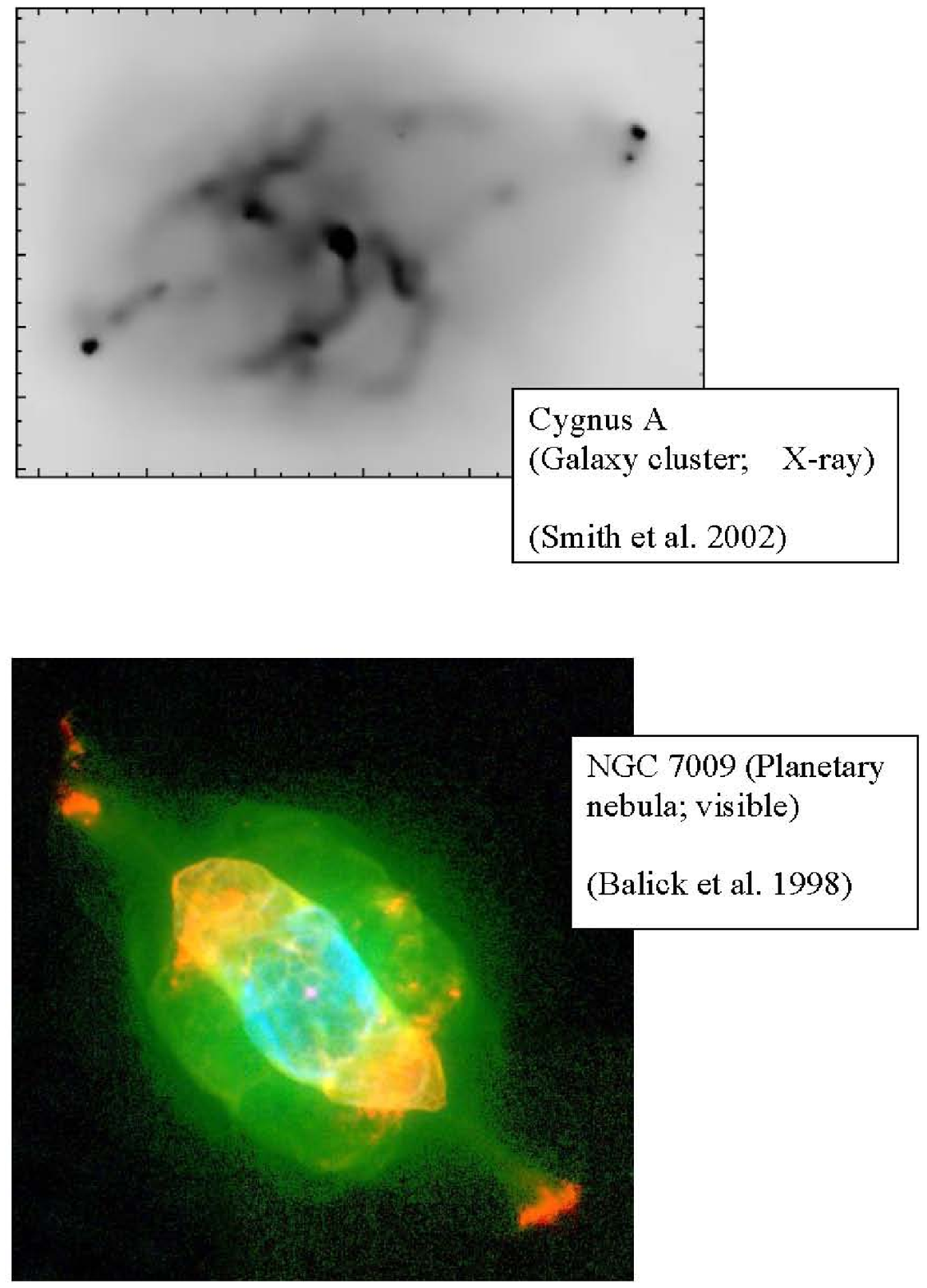} 
\caption{Like Figure \ref{figap1} but emphasizing
pairs of bright bullets along the symmetry axis.}
\end{figure}
\begin{figure}
\includegraphics[width =150mm]{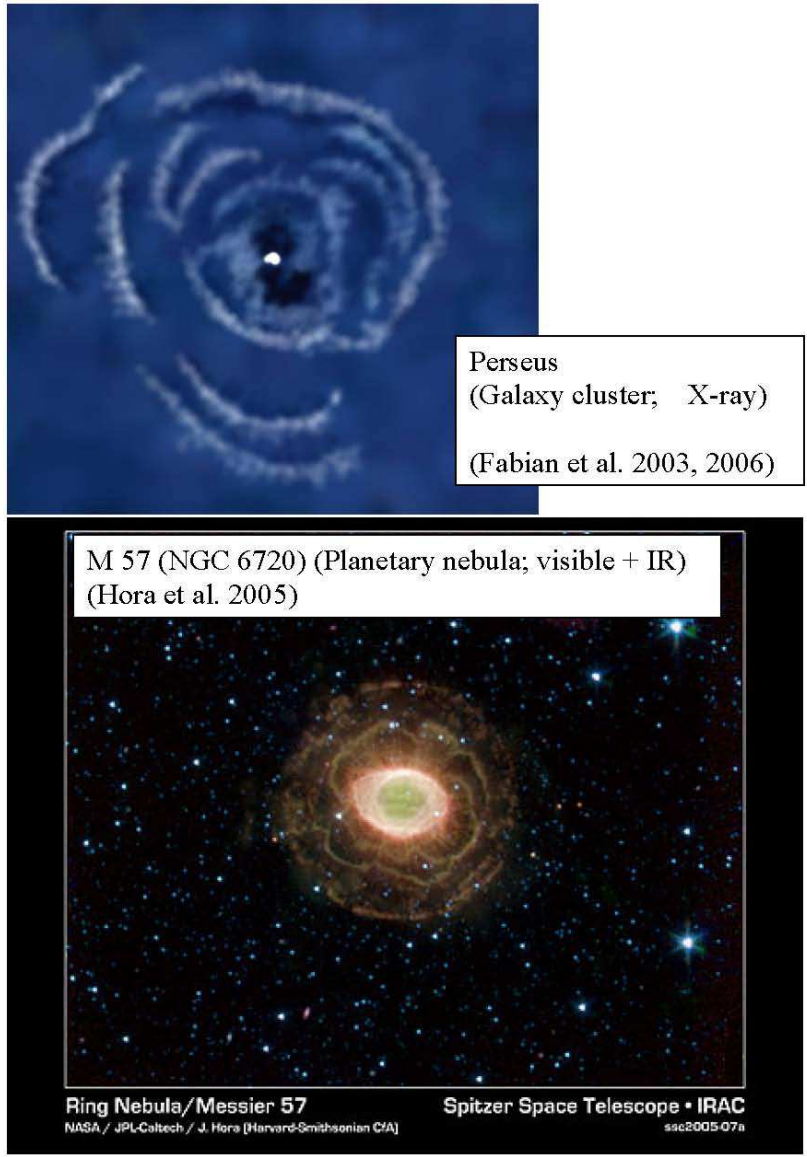}
\caption{Like Figure \ref{figap1} but emphasizing ripples.}
\end{figure}

\end{document}